\documentclass[useAMS,usenatbib]{mn2e}
\usepackage{graphicx}

\newcommand{\gx}{4U~1728--34}
\newcommand{\intg}{\emph{INTEGRAL}}
\newcommand{\rxte}{\emph{RXTE}}
\newcommand{\apj}{ApJ }
\newcommand{\mnras}{MNRAS }

\title[Spectral states evolution of \gx\/ ]{Spectral states evolution of \gx\/ observed by INTEGRAL and RXTE: non-thermal component detection}
\author[A. Tarana]{A. Tarana$^{1}$\thanks{E-mail:antonella.tarana@iasf-roma.inaf.it}, T. Belloni$^{2}$   A. Bazzano$^{1}$, M. M\'endez$^{4}$ and P. Ubertini$^{1}$ \\
$^{1}$Istituto di Astrofisica Spaziale e Fisica Cosmica --INAF, via del Fosso del Cavaliere 100, I-00133 Roma, Italy\\
$^{2}$Osservatorio Astronomico di Brera -- INAF, Via E. Bianchi 46, I-23807 Merate (LC), Italy\\
$^{4}$Kapteyn Astronomical Institute, University of Groningen, PO Box 800, 9700 AV, Groningen, The Netherlands
}
\begin{document}

\date{Accepted...Received...}

\pagerange{\pageref{firstpage}--\pageref{lastpage}} \pubyear{2002}

\maketitle

\label{firstpage}

\begin{abstract}

We report results of a one-year monitoring of the low mass X-ray binary (LMXB) source (atoll type) \gx\/ with \intg\/ and \rxte\/. Three time intervals were covered by \intg\/, during which the source showed strong spectral evolution. We studied the broad-band X-ray spectra in detail by fitting several models in the different sections of the hardness-intensity diagram. The soft states are characterised by prominent blackbody emission plus a contribution from a Comptonized emission. The hard states are characterised by the presence of an excess flux with respect to the Comptonization model above 50 keV while the soft component is fainter.
To obtain an acceptable fit to the data this excess is modeled either with a power law with photon index $\Gamma$ $\sim$ 2 or a Comptonization (CompPS) spectrum implying the presence of hybrid thermal and non-thermal electrons in a corona. This makes \gx\/ one of the few LMXBs of atoll type showing non-thermal emission at high energy. From our analysis, it is also apparent that the presence of the hard tail is more prominent as the overall spectrum becames harder. We discuss also alternative models which can discribe these hard states.


\end{abstract}

\begin{keywords}
X-rays: binaries -- stars: neutron -- individual: \gx\/ 
\end{keywords}

\section{Introduction}

X-ray Binaries show different spectral states depending on the different contribution of the emission components. The neutron star low mass X-ray binary (NS LMXBs) spectra are complex and difficult to explain, and the precise origin of the spectral components are still debated. Different models in fact can describe the same observed spectra. Historically, concerning the soft X-ray range (1-20 keV), two main models were used for the X-ray spectra of LMXBs: the "eastern model" consists of the sum of an optically thick multi-temperature disk, plus a Comptonized blackbody originating from the neutron star or boundary layers between the disk and the neutron star \citep[]{Mitsuda, mitsuda89}; the "western model" uses a Comptonised spectrum due to repeated inverse Compton scattering of the soft seed photons by hot electrons with a thermal distribution of velocities, plus a single temperature blackbody emission \citep{white86}.

In the past years observations also in the hard X-ray bands ($>$20 keV) indicated that broad-band spectral studies are necessary to model and explain all the components and features observed in the soft and hard spectral states of NS LMXBs.

During the soft spectral state the soft thermal component is predominat. The thermal component could come from differents emission regions or combination of these, such as either the optically thick and geometrically thin accretion disk that emits as a black body with a local temperature depending from the disk radius (T(R)$\propto R^{-3/4}$), or the thermal surface of the NS, or the NS boundary layers. Indeed the emission regions of these components are generally known although difficult to discern from the observed spectra. 

During the hard spectral state the soft component decreases or is even not detected, while the contribution of the Comptonization component due to the repeated Compton upscattering of the soft photons in a hot electron cloud (corona) is predominant in the spectra.

The hard component could extend above 200 keV, and in some cases no cut-off at high energy is observed \citep[]{DiSalvo2000a, DiSalvo2001a, iaria2001, piraino1999}, which is the signature for non-thermal emission. The spectral state transition from soft to hard state is often modelled in terms of a gradual increase of the electron temperature of the Comptonising region which is typically in the range of 3-10 keV for the soft states and it increases to the range 10-100 keV for the hard ones. 

Moreover, depending on the geometry of the binary system and on the position of the disk respect to the observer, the additional reflection component could be present in the spectra of the LMXBs. This is due to Compton reflection of the hard X-ray by material of the accretion disk \citep{disalvostella}.

Recently Lin et al. (2007) applied a new model for the high-luminosity soft states of LMXBs with NS, ispired to the black hole binaries (BHBs) behaviour. This model consists of blackbody plus multicolor disk blackbody with no need for a strong Comptonized emission. However, as the luminosity decreases, the Comptonized component is required also for this model.


\gx\/ is a LMXB of the atoll class \citep[]{forman, hasinger, disalvo2001, straaten} containing a NS as indicated by the observation of type-I X-ray bursts \citep[]{lewin, hoffman}. The detection of Eddington-limited bursts allowed a distance estimate of 4.4-5.1 kpc \citep{disalvo, Galloway}.

The X-ray spectrum of the persistent emission of \gx\/ over a wide energy band has been studied with different satellites in the past years: Einstein \citep[]{grindlay81}, \emph{GRANAT} \citep{claret}, \emph{BeppoSAX} \citep[]{disalvo, piraino2000}, \emph{Chandra}, \rxte\/ \citep{d'ai} and \intg\/ \citep{falanga}. The broad-band spectra were modeled with a disk blackbody emission plus a thermal-Comptonization component with varying optical depth and electron temperature \citep[]{disalvo, farinelli, falanga}. Residuals from the fit in the 6-9 keV interval were interpreted either as a broad ($\sigma \sim$ 0.3-0.7) Gaussian emission line at 6.7 keV or as two absorption edges of ionized iron at 7 and 9 keV  \citep{piraino2000, d'ai06}.

\gx\/ is one of the few atoll sources detected in the radio band up to now \citep[]{migliari, marti}. The radio and X-ray fluxes follow a correlation similar to that found for black-hole binaries (BHs) and Z sources which are typically more luminous than atolls \citep[]{migliari06, fender}. Z and BH sources could show a non-thermal tail in the X-ray energy spectra; the presence of this hard tail seems to be correlate with the position of the source in the color-diagram and the radio properties \citep{migliari07}. The analysis of non-thermal spectral components in atoll sources is, in general, more complex than in brighter X-ray binaries due to the low statistics involved, though \intg\/ recently provided important results \citep{tarana1820, fiocchi1636}. The hard tail component could originate from a hybrid thermal/non-thermal population of electrons in a corona near the compact object (\citep{poutanen}) as already revealed in the atoll 4U 1820--30 in its hard state \citep{tarana1820}. Alternatively, the hard tail component could originate from non thermal electrons coming from the base of a jet \citep{markoff}. The bulk motion Comptonization model (BMC) \citep{TitZan} applied usually for BH, was recently used also in NS (CompTB model) and represents an alternative physical explanation of the detected hard tails \citep[]{farinelli07, farinelli08}.


In this work we analyse data from a year-long RXTE/INTEGRAL monitoring campaign of \gx\/ and show a non-thermal tail detected in the X-ray spectra of this source.



\begin{figure}
\includegraphics[width=8.7cm]{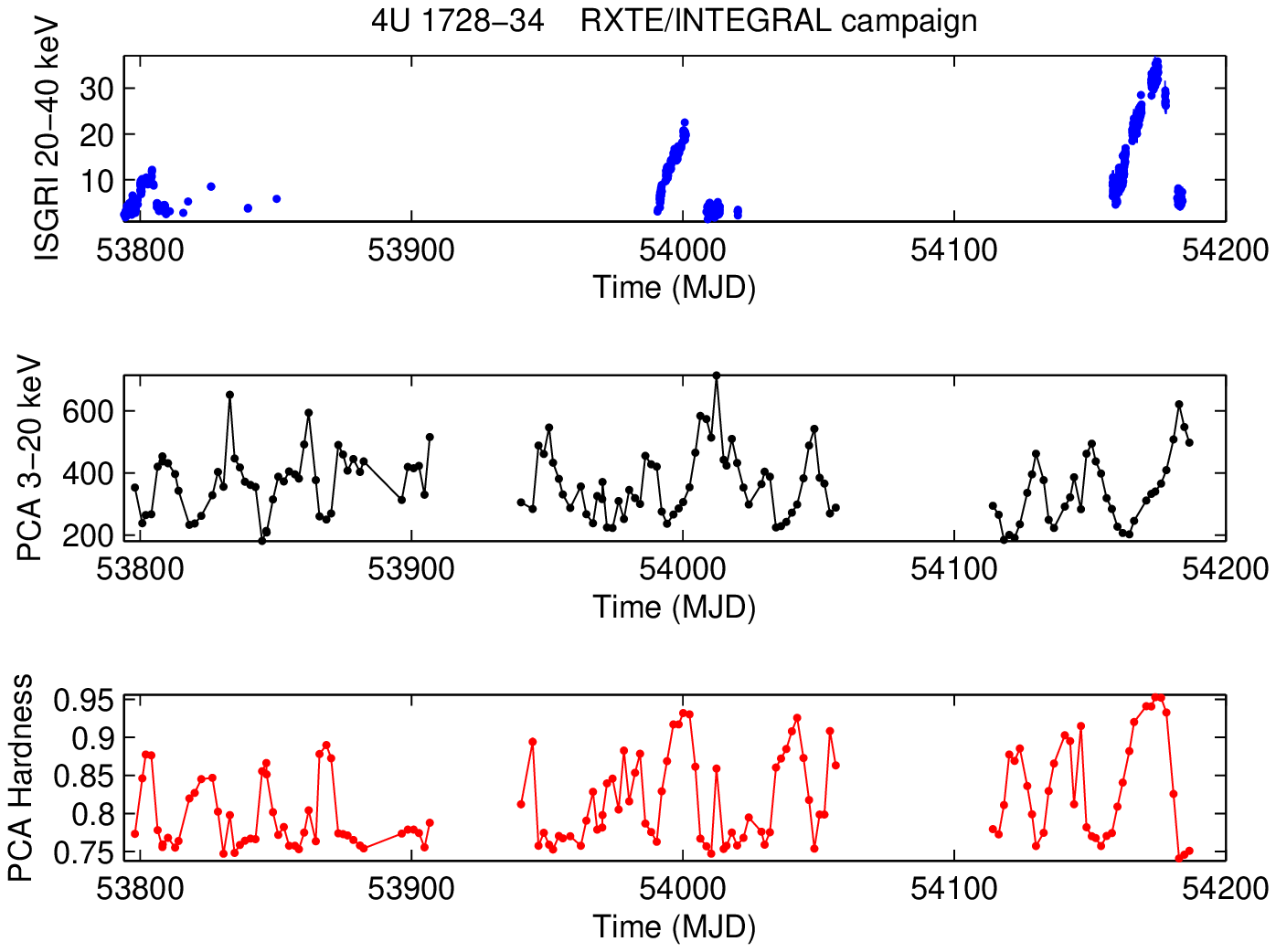}
\includegraphics[width=8.7cm]{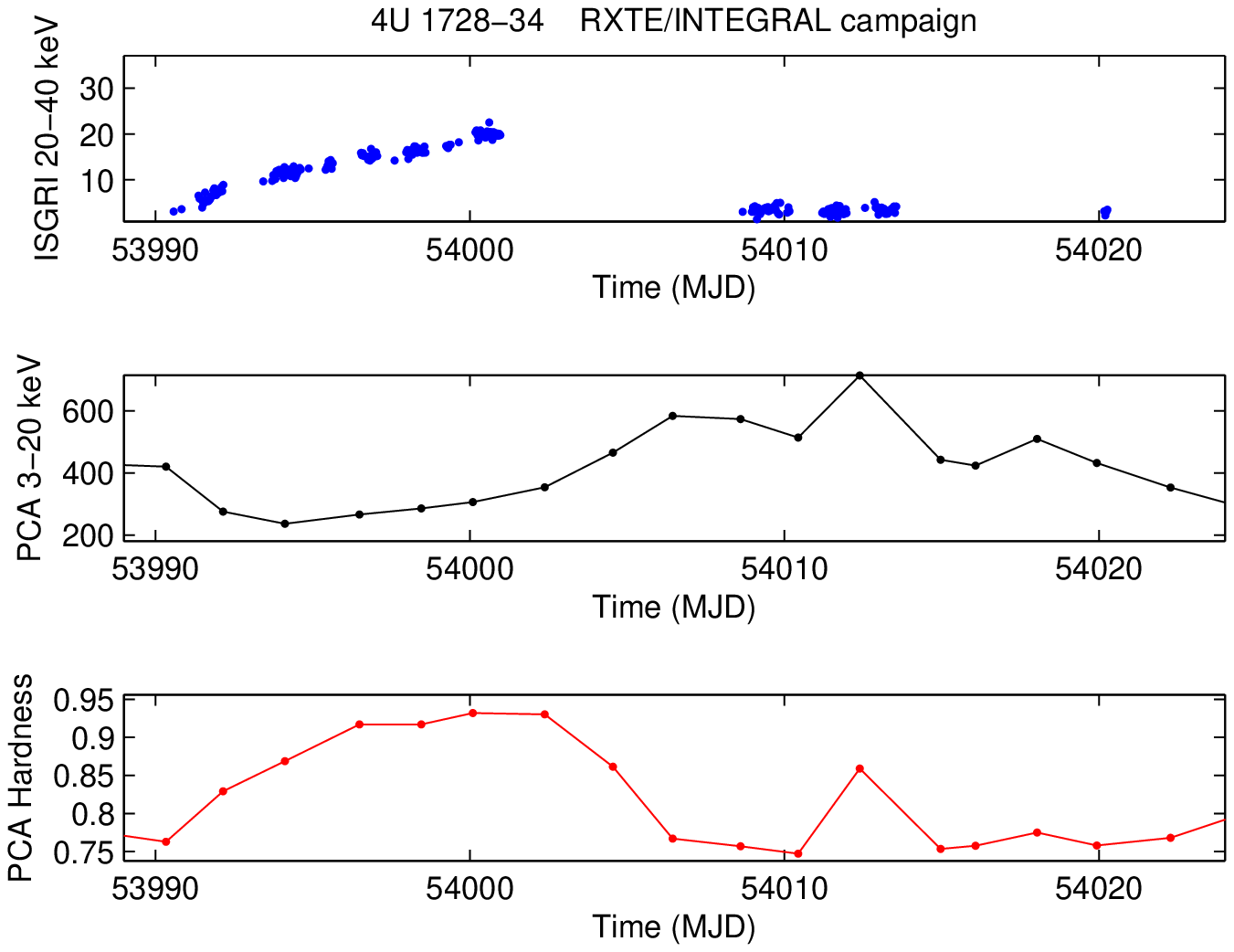}
\caption{\emph{Top:} \intg\//IBIS and \emph{RXTE}/PCA 2006--2007 light curves of  \gx\/. Three main periods of observation are visible. PCA hardness (5.71--9.81 keV/2.87--5.71 keV count rate ratio) is shown in the third panel. \emph{Bottom:} Zoomed-in light curve during the second period.}
\label{licu}
\end{figure}

\begin{figure}
\centering
\includegraphics[width=6.0cm,angle=90]{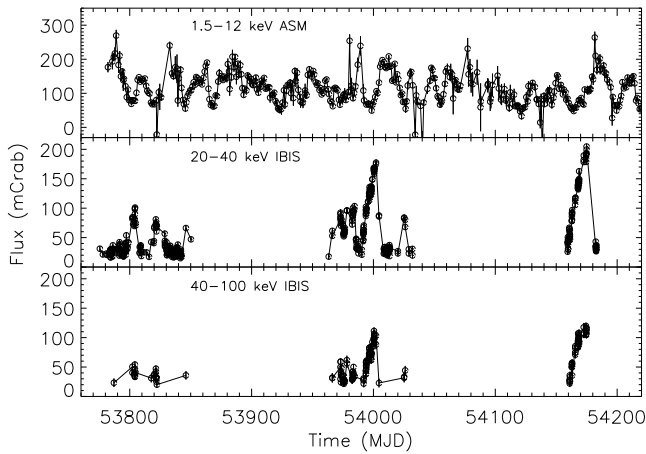}
\caption{\intg\//IBIS and \emph{RXTE}/ASM 2006--2007 light curves of 4U 1728--34. The energy bands are indicated for each panel.}\label{cluce_ibis_GX}
\end{figure}

\begin{figure}
\centering
\includegraphics[height=6.9cm]{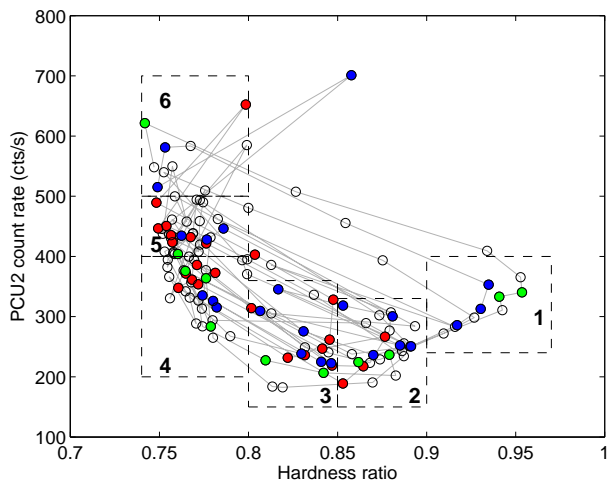}
\caption{Hardness-Intensity diagram of \gx\/ with PCA data. The hardness is the count rate ratio of the bands (5.71--9.81 keV)/(2.87--5.71 keV) and the intensity is the count rate in the 2.87--20.20 keV. Different colors are used for the three different observing periods (see text). The source covers the atoll track moving from the soft (box 6) to the hard (box 1) state, then jumping back to soft.}\label{colorcolorGX}
\end{figure}

\section{The observation campaign}

The \intg\/ monitoring of \gx\/ consists of two parts: the first (February-April 2006) is part of Galactic Centre Deep Exposure observations and the second of Key Program pointings as well as public data (General Program) (2006 August-October and 2007 February-April). The total pointings with our target positioned in the Fully Coded Field of View (FCFOV) are 496 (about 2000 seconds each). We extracted light curves in the 20--40, 40--100 and 100--300 keV energy bands with IBIS. 

\intg\/ data were processed using the Off-Line Scientific Analysis (OSA version 7.0) \citep{gold} software released by the ISDC \citep{curvosier}.

For spectral extraction, all the available data for which the source was within the IBIS/ISGRI FCFOV (4.5$^{\circ}\times$4.5$^{\circ}$) were analysed, so that the flux evaluation is not affected by calibration uncertainties in the off-axis response.

A 2\% systematic error to IBIS/ISGRI data sets was added\footnote{http://isdc.unige.ch/?Support+documents} and the standard 2048 channels response matrix logarithmically rebinned to 62 channels was used.

The \emph{RXTE} monitoring consists of 2 ks pointings every  two days, from 2006 March to 2007 June. We extracted the 3--20 keV light curve from PCA and energy spectra from the PCA and HEXTE instruments (background and deadtime corrected) for each observation using the \rxte\/ software within HEASOFT V. 6.6.3, following the standard procedures. For spectral analysis, only Proportional Counter Unit 2 from the PCA and Cluster B from HEXTE were used. A systematic error of 0.6$\%$ was added to the PCA spectra to account for uncertainties in the instrument calibration. We accumulated background corrected PCU2 rates in the channels A$=$7--48 (2.87--20.20 keV), B$=$7--13 (2.87--5.71 keV) and C$=$14--23 (5.71--9.81 keV). 
The Hardness-Intensity Diagram (HID) was constructed as H $=$ C/B vs. the total rate A (see  \citep{homanbelloni05}). One rate/hardness point was extracted for each observation.

\subsection{Light curves}

Figure \ref{licu} (top three panels) shows the IBIS and PCA light curves of the total observing period in the  20--40  and  3--20 keV  energy bands; the bottom panel shows the time evolution of the PCA hardness ratio corresponding to the count rate ratio 5.71--9.81/2.87--5.71 keV.
In the three intervals with simultaneous  \intg\/--\emph{RXTE} coverage
both flux and hardness show evident changes. A zoom in of the second period is shown in the bottom three panels of Fig. \ref{licu}.

Figure \ref{cluce_ibis_GX} shows the IBIS light curves for the FCFOV pointings in the 20--40, 40--100 keV and in the top panel the 2--12 keV rate of the {\it RXTE}/ASM\footnote{http://xte.mit.edu/ASM$\_$lc.html}. Note that during each \intg\/ observation period a slow monotonic flux increasing in rate followed by a sharp drop is detected.

The anti-correlation of the soft and hard flux indicates a spectral change of the source, as confirmed by its track through the PCA HID of Fig. \ref{colorcolorGX}.

\subsection{The hardness-intensity diagram}

Figure~\ref{colorcolorGX} shows the HID for PCA/\emph{RXTE} data. Different colors corresponds to the three periods for which we have simultaneous data from \emph{RXTE} and \intg\/: red, blue and green points for the first, second and third period respectively. The empty points correspond to  PCA data without \intg\/ coverage. In each period, the source moves steadily from the soft to the hard state and back to the soft state following a counter-clockwise path.
From the HID, we selected six boxes (shown in Fig. 3) and derived energy spectra (hereafter spe1 through spe6) by adding data within each box. In this way, we combine \intg\/ and \emph{RXTE} data corresponding to each different spectral state.
%
%
%

The spectral grouping based on the time evolution of the data, keeping the three observation periods separately, gives spectral changes consistent with the results obtained using the spectra from the selected boxes. We use spectra from the HID boxes which have large signal to noise ratio by adding data of the three periods and by covering a rather large range in count rates. We neverthless checked that the spectra of single observations within each box did not deviate significantly from each other and also that there are no systematic differences between the three periods.

Boxes 1--3 correspond to intervals when the source spectrum was hard (corresponding to the flux maxima of the INTEGRAL light curves), while boxes 4--6 correspond to intervals when the source spectrum was soft.

Note that during the first period 4U 1728-34 did not reach values of hardness as high as in the other periods (no red points are in box 1).



\subsection{Spectral analysis}

We analysed the spe1-spe6 spectral data sets (see Fig.\ref{colorcolorGX}) of the {\it INTEGRAL\/}/IBIS and {\it RXTE\/}/PCA and HEXTE simultaneous observations using XSPEC version 12.5.1 \citep{arnauld}. 


During the fitting a multiplicative constant, frozen to 1 for PCA and a values ranging from 0.7 to 0.9 for IBIS/ISGRI and HEXTE, is added,  to account for uncertainties in the cross calibration of the three instruments.

For the spectral fits we used data in the 3.5--25 keV band for PCA and above 20 keV and 19 keV for HEXTE and IBIS, respectively.

%
%
%


For the different spectral states we attempted several combinations of models. 
To fit a broad emission line we added a Gaussian component that was fixed at the best fit value comprised in $\sigma_{Fe} \sim$ 0.3 -- 0.7 and at about 6.5 keV, compatible with previously observed values  \citep[]{piraino2000, d'ai}.
We fixed the interstellar column density to $N_H =2.3\times10^{22}$ cm$^{-2}$, as deduced from previous observations below 3 keV \citep{d'ai06}. 

We found it difficult to statistically constrain the soft parameters (the blackbody temperature and normalization, and also the seed photons temperatures of the Comptonization component) probably due to the lack of data coverage at low energy ($\la$ 3.5 keV). Hence in some case we fixed the soft parameters to the best-fit value.

\begin{figure}
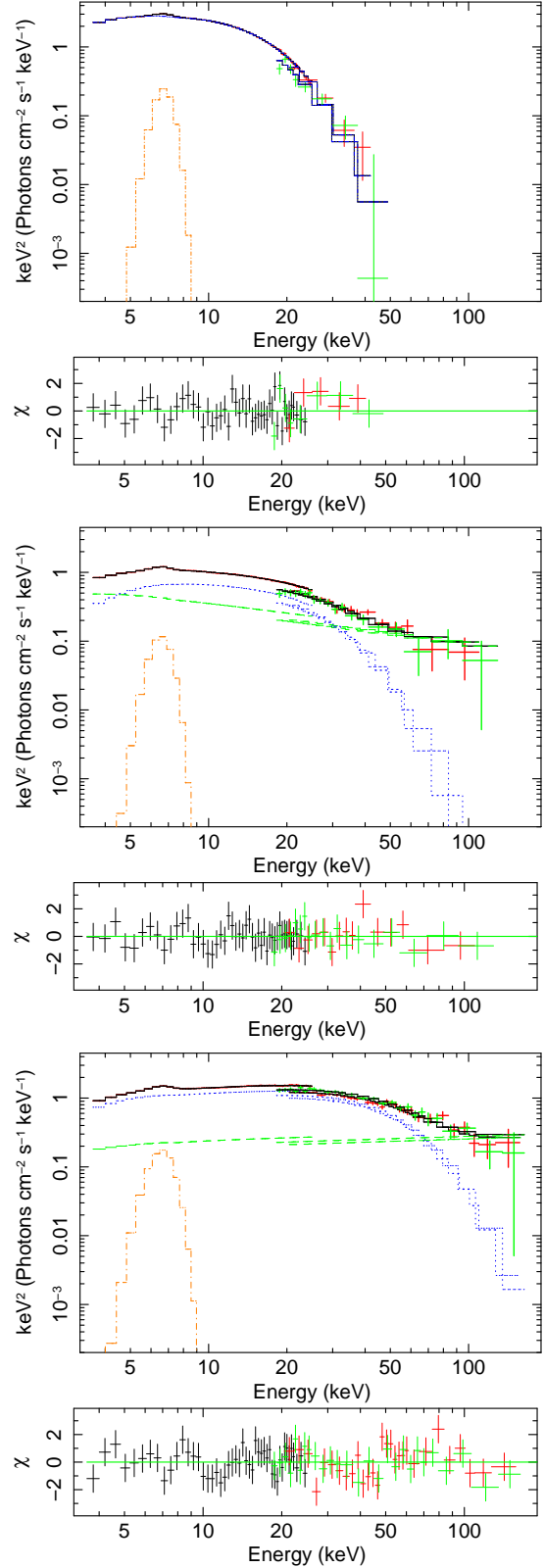

\centering
\includegraphics[height=7.2cm,angle=-90]{spe6_senzabb_eeuf.ps}
\includegraphics[height=7.1cm,angle=-90]{spe6_senzabb_del.ps}
\includegraphics[height=7.2cm,angle=-90]{spe3_senzabb_eeuf.ps}
\includegraphics[height=7.1cm,angle=-90]{spe3_senzabb_del.ps}
\includegraphics[height=7.2cm,angle=-90]{spe1_senzabb.ps}
\includegraphics[height=7.1cm,angle=-90]{spe1_del_color_new.ps}
\caption{\intg\//IBIS (green data) and \emph{RXTE}/PCA and HEXTE (red data) spectra of \gx\/: spe6, spe3 and spe1 data sets from top to bottom, with data, models and residuals to the models with the $\chi$ value. The model is Gaussian plus Comptonization for spe6; and Gaussian, Comptonization plus power law component for spe3 and spe1.}
\label{speGX}
\end{figure}

(1) \texttt{COMPTT} model

\begin{figure}
\centering
\includegraphics[height=7.1cm,angle=-90]{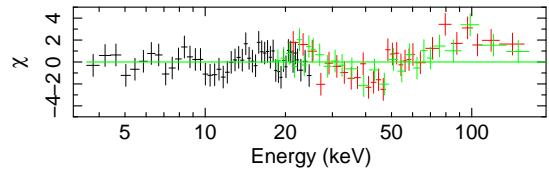}
\caption{Residual to the CompTT models of the spe1 data set, without the power law component.}
\label{residui}
\end{figure}

Only spectra spe5 and spe6 are well represented  with a simple Comptonization model, assumed with a disk geometry, \texttt{CompTT} \citep{sun}, as reported in Table \ref{tab_fitGX}. The results for spectra spe1-spe3 are not acceptable due to an excess in the residuals at high energy see the Fig. \ref{residui}) (at E$>$50 keV, from IBIS and HEXTE). We verified that this excess is not due to the procedure of combining spectra as it is present also in the individual spectra. Moreover, for spe3 and spe4 also a low energy excess (below 5 keV) is detected, probably due to the presence of a soft blackbody emission.

The spectrum spe6  is shown in Fig.~\ref{speGX}. The unabsorbed bolometric flux of the softest spectral state, spe6, is 1.1$\times 10^{-8}$ erg cm$^{-2}$ s$^{-1}$ corresponding to a luminosity of 2.8 $\times 10^{37}$ erg s$^{-1}$, assuming a distance of 4.6 kpc \citep{disalvo, Galloway}. 

(2) \texttt{COMPTT+BB} models

Adding a disk blackbody emission to the Comptonization component, \texttt{diskbb} \citep{Mitsuda} the fits improves for the soft states [the $\chi_r^2$(d.o.f.) is of 1.34(45), 0.85(56) and 0.87(50) respectively for spe4, spe5 and spe6] though this component is not statistically significant because the normalization is not constrained. 
The fits with the addition of a single blackbody (\texttt{bbody}) are better constrained and results are reported in Table \ref{tab_fitGX}. Note that the $\chi_r^2$ is acceptable for all the soft states and also for spe3 even if the residuals to the models still reveal an excess above 50 keV. For spe1 the $\chi_r^2$  is worse than that of a simple Comptonization component and for spe2 the best-fit temperature of the blackbody is very low and difficult to constrain. In conclusion the CompTT+bb model can be applied to the soft states for which  the blackbody temperature decreases and the electron plasma temperature increases from spe6 to spe4. For the hard states: the spectral parameters ($\tau$ and $kT_{\rm bb}$) do not change monotonically from spe3 to spe1 (as the source becomes harder) as in general observed for these type of sources, moreover the  $\chi_r^2$ is not good (see spe1 results). So for hard states it is statistically difficult to constrain the possible soft component together with the Comptonization model.



(3) \texttt{COMPTT+PL} models

To take into account the high energy residuals observed with the simple Comptonization model during the hard state we add a power law component. This component is not required for the soft states. The best-fit parameters of all hard state observations are shown in Table \ref{tab_fitGX}. The power law component becomes less steep as the high energy flux (at $>$ 50 keV) contribution increases.

This is visible in the energy spectra and residuals to the models of spe1 and spe3 which are shown in the Fig. \ref{speGX}. 




(4) \texttt{BB+DISKBB}

For spe6, corresponding to the highest luminous state, a simple double blackbody emission (i.e. \texttt{bb} plus \texttt{diskbb}) gives good fit to the data ($\chi_r^2$ of 0.9) while for the other soft states and for the hard states the fits are not acceptable (the $\chi_r^2$ gives values of 1.8, 2.5 for spe5 and spe4 and become worse for hard states). The blackbody emission obtained for spe6 is higher (2.7 keV) than the value obtained with model (2), and the inner disk temperature is 1.74 keV. 
The high energy emission contribution increases as the source goes to hard states, so other components than the double blackbody are necessary. We try to use two different hard component as described in the following models (4a) and (4b). In both cases to fits the data using the double blackbody it is necessary to freeze the soft component parameters (either the temperature or the normalization values) otherwise the fits with the double blackbody model are not costrained. Moreover the double blackbody is any more necessary for the hard states (in fact only one is sufficient to take into account the soft emission). 


(4a) \texttt{BB+DISKBB+BKNPOWER+HIGHECUT} models

We try a phenomenological model (used in Lin et al. 2007 and motivated by similarities with the BH systems) composed by a two thermal component, a blackbody and a multicolor disk blackbody, plus a broken power law  (\texttt{bknpower}) with high-energy cutoff (\texttt{highecut}).


Results for all the spectra are reported in Table \ref{tab_fitGX}. Freezing the soft parameters (black body normalizations) we obtain fits with a monotonic spectral evolution of the parameters from the soft to the hard states. The blackbody temperature, $kT_{\rm bb}$,  decreases monotonically from the soft to the hard state, while the second soft disk blackbody component is detected only in the softer states.

(4b) \texttt{BB+DISKBB+COMPTT+PL} models

The hard components to add to the double blackbody components could be described as a Comptonization model plus an additional power law that is required for the hard states only. The results are reported in Table \ref{tab_fitGX}. Also in this case we have the same evolution of the soft parameters (black body temperatures) but, in addition, we get the physical information about the origin of the hard component. The electron temperature of the corona increases and the optical depth decreases as the spectrum becomes harder, while the temperature of the seed input photons remains constant.

Even if there is a monotonic physical evolution of the soft parameters (blackbody temperature increases from hard to soft states), the models (4a) and (4b) are not well constrained, in fact to obtain a good fit we need to freeze the  normalizations value of the blackbody.



\begin{figure}
\centering
\includegraphics[height=7.2cm,angle=-90]{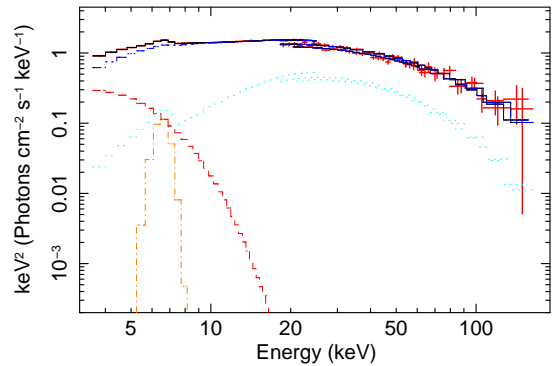}
\caption{Spectrum of the spe1 data set with Gaussian, diskbb plus CompPS model assuming a hybrid thermal and non-thermal electron distribution.}
\label{speCOMPPS}
\end{figure}
\begin{figure}
\centering
\includegraphics[height=7.2cm,angle=-90]{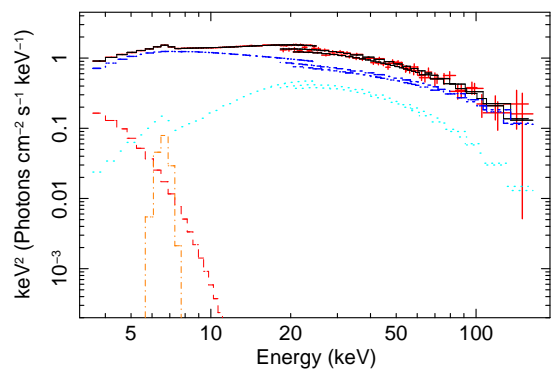}
\caption{Spectrum of the spe1 data set with Gaussian, diskbb, Comptonization plus high reflection component.}
\label{speREFL}
\end{figure}

\begin{table*}\footnotesize
\begin{center}
\caption{Other spectral fitting results for the hard state  spectra of 4U~1728--34 with simultaneous {\it INTEGRAL\/} and {\it RXTE\/} observations}
\label{tab_fitGX_hard}
\begin{tabular}{lccc}
\hline
\hline
                      & spe1 & spe2  &  spe3 \\
\hline
parameters &   \multicolumn{3}{c}{CompPS+diskbb (5)}\\
\hline
$kT_{in}$ (keV)                      & 1.11$^{+0.04}_{-0.14}$     & 1.04$_{-0.05}^{+0.04}$  & 0.93$_{-0.02}^{+0.02}$ \\
norm$_{\rm diskbb}$ $\times 10^{-2}$ & 39 (fx)                    & 47 (fx)                 & 107 (fx) \\ 
$kT_{\rm e}$ (keV)     	             & 26.84$^{+0.78}_{-0.22}$    & 24.20$_{-1.44}^{+1.15}$ & 20$\pm$3.7  \\
$kT_{bb}$ (keV)                      & 1.47$^{+0.08}_{-0.17}$     & 1.46$_{-0.09}^{+0.08}$  & 1.53$_{-0.04}^{+0.05}$ \\
$\tau_{y}$        		     & 3.0$\pm$1.24               & 2.83$\pm$0.31           & 2.43$_{-0.18}^{+0.06}$ \\
$\Omega/2\pi$ 			     & 1.0 $\pm$0.43              & 1$\pm$0.37              & 1$\pm$0.57\\
norm$_{\rm CompPS}$                  & 55.36$^{+38.82}_{-11.53}$  & 42.32$_{-8.80}^{+13.36}$  & 35.83$_{-4.20}^{+4.60}$\\ 
$\chi_r^2$($\nu$)                    & 0.77(95)                   &  0.78(84)               & 0.69(75)        \\
\hline
parameters &   \multicolumn{3}{c}{CompTT+diskbb+reflect (6)}\\
\hline
$\Omega/2\pi$ 			     & 1.27$^{+0.38}_{-0.27}$     & 1.20$^{+0.40}_{-0.22}$ & 1.51$^{+0.40}_{-0.36}$\\ 
$kT_{0}$ (keV)        	             & 1.14$^{+0.04}_{-0.05}$     & 1.19$_{-0.03}^{+0.02}$ & 1.45$^{+0.41}_{-0.04}$  \\
$kT_{\rm e}$ (keV)   		     & 30.66$^{+26.03}_{-8.24}$   & 28.82$\pm$13.50        & 28.47$\pm$14.90   \\ 
$\tau$                               & 0.76$^{+0.42}_{-0.46}$     & 0.70$^{+0.38}_{-0.02}$ & 0.39$\pm$0.33   \\ 
norm$_{\rm CompTT}$$\times 10^{-2}$  & 1.63$^{+0.74}_{-0.82}$     & 1.27$_{-0.02}^{+0.18}$ & 0.97$_{-0.68}^{+0.85}$ \\
$kT_{in}$ (keV)                      & 0.73$^{+0.03}_{-0.04}$     & 0.77$_{-0.03}^{+0.01}$ & 0.95$_{-0.01}^{+0.01}$ \\
norm$_{\rm diskbb}$ $\times 10^{-2}$ & 236 (fx)                   & 191 (fx)               & 115 (fx) \\ 
$\chi_r^2$($\nu$)                    & 0.74(95)                   &  0.78(84)              & 0.71(75) \\
\hline
\hline
\end{tabular}
\end{center}
\end{table*}

\subsubsection{Further models for the hard states}

The presence of high energy residual to the Comptonization model for the hard states, spe1-spe3, prompted us try also other models.

(5) \texttt{DISKBB+COMPPS} models

The \texttt{CompPS} model \citep{PouSven}, which takes into account the Comptonized processes with non-thermal distribution of the velocities of the electrons, was used to model some binary systems showing non-thermal component. So we combine a disk blackbody component and a \texttt{CompPS} modelled with the assumption of a hybrid thermal and non-thermal composition of the electrons plasma in a corona, which takes into account the high energy excess, and a good fit has been obtained. The result is a Comptonizing plasma, assumed of spherical geometry, with $kT_{\rm e}$ $\simeq$ 20--27 keV, optical depth $\tau_{y}$ $\simeq$ 2.4--3, input seed photons with $kT_{bb}$ $\simeq$ 1.5 keV, reflection factor, $\Omega/2\pi$, of 1, inner disk temperature of 1 keV, and the high energy tail with electron power law index $p$ $\simeq$ 0.6, 1.37 and 3.2 respectively for spe3, spe2 and spe1 above the electron Lorentz factor of 1.6. Results are reported in Table \ref{tab_fitGX_hard}. 
 We show the energy spectra with these models for the hardest spectrum (spe1) in Fig. \ref{speCOMPPS}. 

The unabsorbed bolometric flux of the spe1 and spe3 spectra is  8.5$\times 10^{-9}$ erg cm$^{-2}$ s$^{-1}$  and 5.8$\times 10^{-9}$ erg cm$^{-2}$ s$^{-1}$ corresponding to a bolometric luminosity of 2.1$\times 10^{37}$ erg s$^{-1}$ and  1.5$\times 10^{37}$ erg s$^{-1}$.

(6) \texttt{DISKBB+COMPTT+REFLECT} models

An alternative model for the hard states consist in adding a reflection component, \texttt{reflect} \citep{refl}, to Comptonization and blackbody. In this case the Comptonization component requires input seed photons with a temperature of 1.1--1.5 keV a corona with a higher electron temperature, ranging from 28 to 31 keV, and a lower optical depth of 0.4--0.8 with respect to the models (3). The inner disk blackbody temperature is 0.7--0.9 keV and the reflection component, which fits the bump near 20--30 keV, has a high reflection factor of 1.2--1.5. We show the energy spectra of the hardest spectrum (spe1) with these alternative models in Fig. \ref{speREFL}. This model gives acceptable $\chi_r^2$($\nu$) as reported in Table \ref{tab_fitGX_hard}. 

This model gives a high reflection fraction which would imply a small inner disc radius not usually observed in the hard state and that makes this model difficult to consider (for a discussion of inner radii of accretion disk in hard state see e.g. \citep{reis} and also \citep{done06}). Moreover, no indication of such strong reflection feature was observed by {\emph{CHANDRA}} (even if below 10 keV) for this source and in general this high value has not been detected so far in the atoll class. 
%

\begin{table*}\footnotesize
\begin{minipage}{\textwidth}
\begin{center}
\caption{Spectral fitting results for the spectra of 4U~1728--34 with simultaneous {\it INTEGRAL\/} and {\it RXTE\/} observations}
\label{tab_fitGX}
\begin{tabular}{l|cccccc}
\hline
\hline
                       & spe1 & spe2  &  spe3 & spe4 & spe5 & spe6  \\
\hline
parameters                 &    \multicolumn{5}{c}{CompTT  (1)} \\
\hline
$kT_{0}$ (keV)             & 0.85$^{+0.02}_{-0.02}$            &    0.87$^{+0.02}_{-0.02}$ & 0.95$\pm$0.01 & 0.82$_{-0.04}^{+0.05}$ & 0.50$\pm$0.06          & 0.72$_{-0.04}^{+0.05}$ \\
$kT_{\rm e}$ (keV)         & 12.3$^{+0.39}_{-0.37}$            &    11.99$_{-0.69}^{+0.54}$& 7.33$\pm$0.23 & 3.18$_{-0.05}^{+0.05}$ & 3.10$_{-0.02}^{+0.05}$ & 3.02$_{-0.03}^{+0.03}$ \\
$\tau$                     & 2.62$^{+0.06}_{-0.06}$            &    2.37$_{-0.07}^{+0.12}$ & 2.81$\pm$0.08 & 5.44$_{-0.15}^{+0.14}$ & 5.75$_{-0.06}^{+0.05}$ & 5.73$_{-0.13}^{+0.09}$ \\
norm$_{\rm CompTT}$  $\times 10^{-2}$ & 5.84$^{+0.21}_{-0.21}$ & 4.61$^{+0.34}_{-0.26}$    & 7.65$\pm$0.28 & 29.65$_{-1.24}^{+1.56}$ & 53.55$_{-2.79}^{+3.89}$ & 59.81$_{-2.85}^{+2.75}$ \\
$\chi_r^2$($\nu$)          & 1.51(98)                          & 1.45(88)                  & 2.32(73)      & 1.95(47)              &  1.17(58)              & 0.88(52)               \\
\hline
             &               \multicolumn{5}{c}{CompTT+bb  (2)}  \\
\hline
$kT_{0}$ (keV)             & 0.93$^{+0.07}_{-0.06}$    & 0.89$_{-0.03}^{+0.04}$  &  0.76$_{-0.05}^{+0.04}$   &  0.70$^{+0.04}_{-0.04}$  & 0.69$^{+0.05}_{-0.25}$ & 0.79$_{-0.05}^{+0.04}$ \\
$kT_{\rm e}$ (keV)         & 12.17$_{-0.41}^{+0.44}$   & 11.96$_{-0.57}^{+0.61}$ &  9.31$_{-0.95}^{+1.37}$   &  6.67$^{+1.51}_{-1.25}$  & 5.45$_{-0.77}^{+0.33}$   & 4.13$_{-0.13}^{+0.05}$  \\
$\tau$                     & 2.64$_{-0.08}^{+0.08}$    & 2.37$_{-0.07}^{+0.09}$  &  2.41$_{-0.30}^{+0.25}$   &  2.26$^{+0.52}_{-0.46}$  & 2.63$_{-0.19}^{+0.15}$   & 3.28$_{-0.14}^{+0.18}$  \\
norm$_{\rm CompTT}$ $\times 10^{-2}$ & 5.46$^{+0.31}_{-0.32}$ &4.49$_{-0.28}^{+0.15}$  &  6.24$_{-0.89}^{+0.82}$  &  12.53$^{+3.49}_{-2.77}$ & 19.48$_{-3.45}^{+14.6}$   & 34.46 $_{-2.84}^{+3.71}$\\
$kT_{bb}$ (keV)            & 0.89$^{+0.18}_{-0.23}$    & 0.30$_{-0.29}^{+0.40}$  &  2.48$_{-0.19}^{+0.22}$   &  2.34$^{+0.42}_{-0.43}$  & 2.39$_{-0.04}^{+0.04}$   & 2.44$\pm$0.05  \\
norm$_{\rm bb}$  $\times 10^{-3}$& 1.75$^{+0.6}_{-0.4}$& 35.89$_{-2.83}^{+1.55}$ &  4.50$_{-0.55}^{+0.53}$   &  19.18$_{-0.87}^{+3.12}$ & 23.25$_{-0.61}^{+0.78}$  & 29.93 $_{-0.64}^{+85.31}$ \\
$\chi_r^2$($\nu$)          & 1.61(96)                  & 1.39(82)                &  0.99(71)                 &  0.82(56)                & 0.94(56)                & 0.91(50)            \\
\hline
                &              \multicolumn{5}{c}{CompTT+PL  (3)} \\
\hline
$kT_{0}$ (keV)             & 0.85$^{+0.08}_{-0.04}$        & 0.89$_{-0.04}^{+0.09}$ & 1.17$_{-0.06}^{+0.07}$ & --  & --  & --\\
$kT_{\rm e}$ (keV)         & 10.05$^{+0.59}_{-0.60}$       & 9.14$_{-0.67}^{+0.80}$ & 5.75$_{-0.19}^{+0.36}$ & --  & --  & --\\
$\tau$                     & 2.99$^{+0.25}_{-0.15}$        & 2.84$_{-0.19}^{+0.25}$ & 3.49$_{-0.15}^{+0.12}$ & --  & --  & --\\
norm$_{\rm CompTT}$ $\times 10^{-2}$ & 5.93$^{+0.8}_{-1.1}$& 4.84$_{-1.05}^{+1.04}$ & 5.19$_{-0.63}^{+0.59}$ & --  & --  & --\\ 
$\Gamma$                   & 1.83$^{+0.22}_{-0.55}$        & 1.92$_{-0.04}^{+0.09}$ & 2.48$_{-0.07}^{+0.09}$ & --  & --  & --\\
norm$_{\rm PL}$            & 0.15$^{+0.28}_{-0.14}$        & 0.17$_{-0.15}^{+0.30}$ & 0.99$_{-0.15}^{+0.21}$ & --  & --  & --\\
$\chi_r^2$($\nu$)          & 0.88(96)                      &  0.89(84)              & 0.60(71)               & --  & --  & --\\
\hline
             &               \multicolumn{5}{c}{bb+diskbb+bknpl+higheut  (4a)}  \\
\hline
$\Gamma_{\rm 1}$           & 1.39$^{+0.08}_{-0.10}$       &1.59$_{-0.08}^{+0.09}$  & 1.90$_{-0.06}^{+0.07}$  & 2.08$\pm$0.16 & 2.44$\pm$0.07  & 2.50$\pm$0.61\\
$E_{\rm break}$            & 14.15$^{+0.60}_{-0.61}$      &13.96$_{-0.87}^{+1.05}$ & 14$_{-0.72}^{+0.77}$   & 18(fx)        & 18(fx)         & 18(fx)       \\
$\Gamma_{\rm 2}$           & 1.89$^{+0.05}_{-0.05}$       &2.04$_{-0.09}^{+0.10}$  & 2.48$_{-0.11}^{+0.11}$ & 3.34$\pm$0.17 & 3.03$\pm$0.22  & 3.81$\pm$0.82\\
norm$_{\rm bknpl}$         & 0.29$^{+0.07}_{-0.07}$       &0.35$_{-0.07}^{+0.08}$  & 0.64$_{-0.09}^{+0.11}$ & 0.29$\pm$0.22 & 0.69$\pm$0.13  & 0.45$\pm$0.30\\
$E_{\rm cutoff}$           & 22.13$^{+1.74}_{-1.30}$      &21.12$_{-0.67}^{+0.80}$ & 19.44$_{-1.34}^{+1.41}$&    --         & --             & --           \\
$E_{\rm fold}$             & 48.70$^{+3.97}_{-3.70}$      &48.23$_{-1.74}^{+3.62}$ & 38.52$_{-5.76}^{+8.80}$&    --         & --             & --           \\
$kT_{bb}$ (keV)            & 1.43$^{+0.02}_{-0.02}$       &1.46$_{-0.04}^{+0.05}$  & 1.59$_{-0.06}^{+0.08}$ & 2.42$\pm$0.01 &  2.54$\pm$0.01 & 2.71$\pm$0.02\\
norm$_{\rm bb}$& 24.9$^{+5.32}_{-4.81}$ &16.72$_{-4.22}^{+4.65}$ & 11.47$_{-3.11}^{+3.37}$& 6(fx)         &  6(fx)         & 6(fx)        \\
$kT_{in}$ (keV)            &     --                       &    --                  &    --                  & 1.47$\pm$0.02 &  1.54$\pm$0.01 & 1.78$\pm$0.01\\
norm$_{\rm diskbb}$        &     --                       &    --                  &    --                  & 30(fx)        &  30(fx)        & 30(fx)       \\
$\chi_r^2$($\nu$)          & 0.62(92)                     & 0.75(81)               &  0.49(70)              & 0.79(48)       & 0.75(59)      & 0.85(53)     \\
\hline
             &               \multicolumn{5}{c}{CompTT+bb+diskbb+PL  (4b)}  \\
\hline
$kT_{0}$ (keV)             &       0.89$\pm$0.08      &1.0(fx)        & 0.95$\pm$0.12  &  1.0(fx)        &  1.0(fx)        & -- \\
$kT_{\rm e}$ (keV)         &       10.47$\pm$0.53     &8.76$\pm$0.38  & 6.55$\pm$1.57  &  5.46$\pm$0.69  &  4.99$\pm$0.56  & -- \\
$\tau$                     &       2.87$\pm$0.10      &2.97$\pm$0.10  & 3.12$\pm$0.71  &  3.21$\pm$0.77  &  3.59$\pm$0.60  & -- \\
norm$_{\rm CompTT}$ $\times 10^{-2}$ & 5.89$\pm$0.37  &4.44$\pm$0.19  & 6.80$\pm$1.29  &  6.77$\pm$2.35  &  6.52$\pm$1.23  & -- \\
$kT_{bb}$ (keV)            &       0.68$\pm$0.34      &0.96$\pm$0.06  & 2.09$\pm$0.41  &  2.22$\pm$0.06  &  2.34$\pm$0.03  & 2.76$\pm$0.01  \\
norm$_{\rm bb}$& 21.44 (fx)         &16(fx)         & 1.1(fx)        &  6.64$\pm$0.55  &  7.5(fx)        & 5.96$\pm$0.039         \\
$kT_{in}$ (keV)            &     --                   &  --           & 0.49$\pm$0.09  &  0.93$\pm$0.04  &  1.13$\pm$0.17  & 1.74$\pm$0.04  \\
norm$_{\rm diskbb}$        &     --                   &  --           & 1575(fx)       &  163(fx)        &  98(fx)         & 37.11$\pm$1.68         \\
$\Gamma$                   &      1.72$\pm$0.01       &1.9$\pm$0.07   & 2$\pm$1.1      &  --             &    --           &  --             \\
norm$_{\rm PL}$            &      0.08$\pm$0.06       &0.16$\pm$0.06  & 0.11$\pm$0.57  &  --             &    --           &  --             \\
$\chi_r^2$($\nu$)          & 1.09(94)                 &0.84(84)      &  0.74(71)       &  0.88(47)       & 0.76(59)        & 0.87(54)        \\
\hline
\hline
F$_{\rm bol}$ (erg s$^{-1}$ cm$^{-2}$)\footnote{The fluxes are extracted by the CompPS plus diskbb models for the spe1-spe3 data and by model (4b) for the spe4-spe6 data} & 8.5$\times 10^{-9}$  & 6.6$\times 10^{-9}$  & 5.8$\times 10^{-9}$   & 7.1$\times 10^{-9}$  & 8.2$\times 10^{-9}$  & 1.1$\times 10^{-8}$  \\
\hline
\end{tabular}
\end{center}
\end{minipage}
\end{table*}

\section{Discussion and conclusions}

The simultaneous {\it RXTE} and {\it INTEGRAL} long monitoring observations allow us to study over a broad energy band (3--200 keV) the different spectral states of \gx\/ selected according to the position on the PCA Hardness-Intensity diagram, as well by the {\it INTEGRAL} high energy behavior of the source.

In general the observed spectra of \gx\/ turned out to be difficult to model.  In particular at low energy the presence of the blackbody or disk blackbody emission are difficult to constrain because the PCA is sensitive for E$>$ 4 keV, and at high energy there seems to be a non-thermal component during the hard state that needs to be investigated. 

A number of models provide statistically satisfactory fits to our data. 
We prefer models that provide spectral parameters with physical meaning and at the same time evolve regularly through the source states.


The simplest spectral model that describes well the observed spectral states is the sum of blackbody and Comptonization (CompTT) emission for the soft states and Comptonization with the addition of the power law for the hard states.

In the soft spectra the Comptonization would arise from a corona with a low electron temperature of about 4--7 keV and high optical depth of 2--3, with input seed photons of temperature 0.6--0.8 keV. The blackbody temperature $kT_{\rm bb}$ is 2.3--2.5 keV and consistently  higher than the seed photons temperature which gives rise to the Comptonization processes. This indicates that the blackbody emission could come either from inner parts of the disk or from the boundary layer while the seed photons could be due to the emission from the outer part of the disk or from the NS surfarce directly. The compatibility of the boundary layer temperature predicted from the spreading layer model \citep{SulPou} with the obtained $kT_{\rm bb}$ supports the hypothesis that the blackbody component originates from the boundary layers.

Only the most luminous soft state could be also modelled with a double thermal emission (simple blackbody plus disk blackbody emission). In this last case the temperature of the blackbody is still compatible with the boundary layer emission while the disk emission has lower temperature (1.7 keV) and would come from the outer part of the disk.
%

The hard spectra can be described by a thermal Comptonization with a temperature of the electrons of $\sim$ 6--10 keV and optical depth of 3, plus a power law with photon index of $\sim$ 1.8--2.5. As the spectrum becomes harder, the temperature of the electrons increases and the optical depth decreases. The seed photon temperature does not change much during the soft to hard spectral variation indicating the same emission region (in this case the hypothesis of the NS emission origin could be suitable).

By combing the results of these two models for the hard and soft states, we note that the variation of the physical parameters values is quite monotonic (within the error values). During the hardening the blackbody temperature, $kT_{\rm bb}$,  decreased (until it is no longer detected in the hardest states), the Comptonization component increased in temperature, $kT_{\rm e}$, and decreased in optical depth, $\tau$, while the seed photons temperature value, $kT_{\rm 0}$, did not vary significantly.


The detection of a hard tail component for this atoll source, even if not the emission prominent in our data, represents an important results, but our data do not allow us to establish a proper origin. The detection of radio emission during the hard state could indicate the link of this non-thermal emission with a jet formation \citep{migliari07}. In fact, as the BHCs show hard states associated with jet formation, this could be also the case for atoll sources. Radio emission was already observed from \gx\/ but unfortunately during our observations no radio coverage was available for this source.

Alternatively, the hard states could be also modelled by a blackbody component with temperature of 1 keV plus a Comptonization (CompPS) with higher electron temperature (20--26 keV) and a high energy tail due to a hybrid thermal-non thermal composition of the electron plasma. Hence the hybrid thermal-non thermal composition of the electron population in the corona appears as a plausible explanation for the hard tails detected. 

We found that the hard states could also be modelled without the use of the Comptonization component but with the use of the phenomenological model consisting of a blackbody plus a broken power law with a high energy cutoff as already used by Lin et al. 2007.

In the hardest spectrum (spe1) the relative flux contribution above E$>$60 keV is about 10$\%$, i.e. similar to the value in the hard state of 4U~1820--30 and, in general, is of the same order of other hard tail flux contribution in NS systems \citep{tarana1820} \citep{migliari07}.

Finally, we note that this source presents flux variations that not always reach the same value of hardness, as shown by the hardness intensity diagram. This indicates that the hardest spectrum characterised by the power law component can not always be reached by the source, this could explain the difficulty to detect this component in previous observations of the source. This could be due to an additional parameter, hitherto unknown, driving spectral variation besides the accretion rate \citep{Homan2001}.

\section*{Acknowledgments}

This research has made use of data obtained with {\it INTEGRAL\/} which is an ESA project with instruments and science data centre funded by ESA member states especially the PI countries: Denmark, France, Germany, Italy, Switzerland, Spain, Czech Republic and Poland, and with the participation of Russia and the USA. The authors thanks M. Federici for the continuous effort to update the {\it INTEGRAL\/} archive and software in Rome and L. Natalucci for the scientific and data analysis support. A special thanks is for J. Homan who gives important scientific contribution to the paper and in particular to the spectral fitting procedure. The authors thanks also A. Paizis for the information and discussion about the applicability of the compTB model to the hard spectral state. The research leading to these results has received funding from the European Community’s Seventh Framework Programme (FP7/2007-2013) under grant agreement number ITN 215212 “Black Hole Universe”. We acknowledge the ASI financial support via grant ASI-INAF I/008/07.


\bsp

\label{lastpage}

\end{document}